\documentclass[10pt,aps,prl,twocolumn,superscriptaddress,floatfix,longbibliography]{revtex4-2}
\usepackage{amsfonts}
\usepackage{siunitx}
\usepackage{amsmath}
\usepackage{amssymb}
\usepackage{graphicx}
\usepackage[colorlinks]{hyperref}
\usepackage{braket}
\usepackage{xcolor}
\definecolor{linkcolor}{RGB}{0,83,166}
\hypersetup{
  colorlinks = true,
  allcolors = {linkcolor}
}

\begin{document}

\title{Moir\'e Control of Alterelectric Quadrupolar Order}

\author{Alejandro Lopez-Bezanilla}
\email[]{alejandrolb@gmail.com}
\affiliation{Theoretical Division, Los Alamos National Laboratory, Los Alamos, New Mexico 87545, USA}
\date{\today}

\begin{abstract}
Alterelectricity is a compensated ferroic state in which quadrupolar electronic order reshapes low-energy electronic structure without producing a net polarization. Here we show that a moir\'e superlattice can turn such order into a controllable phase. Within a Bloch-periodic two-orbital theory, the slowly varying interlayer registry is coarse-grained into an effective moir\'e field acting on a self-consistent two-component alterelectric quadrupole. The resulting phase develops above a strongly filling-dependent instability threshold and crosses over from a weakly selected regime into a robust axial-dominated ground state, while the diagonal-dominated branch remains only a weak competitor. A registry-phase sweep supplies an explicit continuous path through internal quadrupole space, demonstrating that the moir\'e superlattice does more than stabilize alterelectricity: it steers its internal orientation. This orientational selection is encoded directly in the redistribution of low-energy spectral weight across the moir\'e Brillouin zone. These results identify moir\'e superlattices as a generic route to controllable alterelectric order and to programmable anisotropic electronic functionality.
\end{abstract}

\maketitle

Compensated broken-symmetry phases that nevertheless leave symmetry-resolved fingerprints in spectroscopy have become a central theme in contemporary quantum materials~\cite{Bohmer2022}. Altermagnetism is the clearest recent example: despite vanishing net magnetization, it produces momentum-dependent spin splitting and has rapidly progressed from symmetry classification to microscopic mechanisms, material searches, and switchable ferroic extensions.~\cite{Smejkal2022PRXClass,Mazin2022Editorial,Radaelli2024Tensorial,Leeb2024PRL,Jungwirth2025,Gu2025,Chen2025} This development motivates a broader extension of compensated-order physics to the charge-orbital sector.  
The recently proposed concept of alterelectricity suggests that such a symmetry-allowed and experimentally tunable realization is indeed possible.~\cite{Alterelectricity2026} An alterelectric (AE) state is characterized by a switchable anisotropy of the electronic structure without a macroscopic electric dipole, implying that the natural local order parameter is quadrupolar rather than polar.~\cite{Alterelectricity2026,Kuramoto2009Multipoles} The first material proposals already emphasize structurally tunable settings  indicating that structural registry may be as central to alterelectricity as sublattice symmetry is to altermagnetism.~\cite{Alterelectricity2026}

Moir\'e superlattices are especially appealing in this context because they convert local stacking into a long-wavelength anisotropic field while preserving direct access to carrier density, local spectroscopy, and relative registry.~\cite{Carr2020Methods,Balents2020MoireFlatBands,Nuckolls2024Moire} This is particularly timely for orbital-active layered systems: moir\'e engineering has matured from a band-structure motif into a platform where charge, excitonic, and orbital degrees of freedom can be selected and interrogated locally.~\cite{Hattori2024OrbitalMoire,Nuckolls2024Moire} The central issue is therefore whether a moir\'e environment can support alterelectric order and provide controlled access to its internal quadrupolar manifold. 

In this Letter, we show that a moir\'e superlattice provides direct control over alterelectric order by acting on its internal quadrupolar orientation. Within a coarse-grained two-orbital model on a periodic moir\'e supercell, we find a strongly filling-dependent crossover into a large-amplitude compensated quadrupolar state. At fixed registry, the square moir\'e field lifts the near-degeneracy between axial and diagonal sectors and selects an axial-dominated ground state over most of the ordered regime. Changing the registry phase then tunes the effective quadrupolar selector itself, producing a continuous path through internal quadrupole space that transfers weight from the axial to the diagonal channel without suppressing the order. The corresponding sector is resolved independently in the momentum-resolved spectral function, providing a direct one-particle fingerprint of alterelectric orientation.

\begin{figure*}
    \centering
    \includegraphics[width=0.99\linewidth]{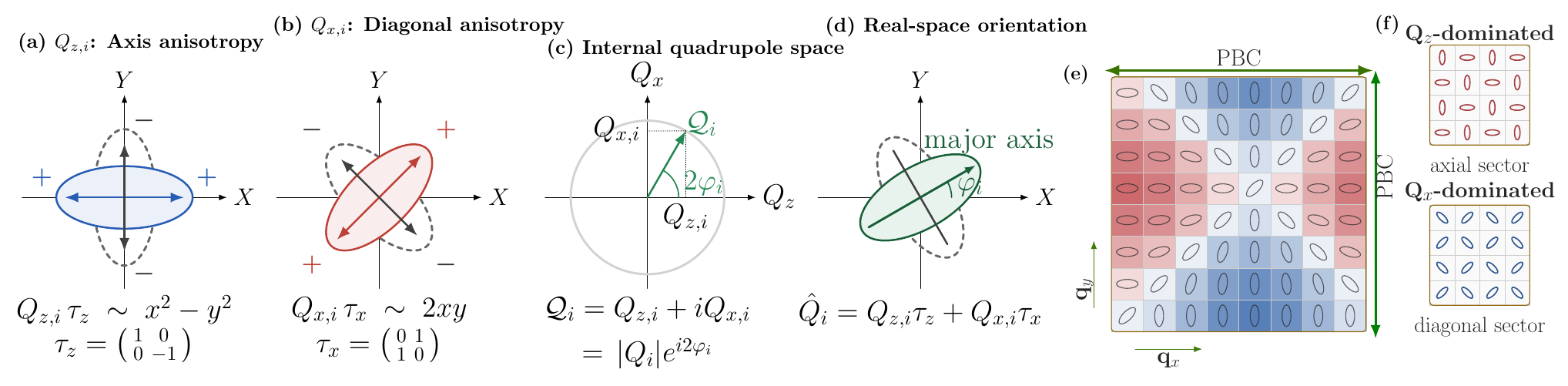}
    \caption{
Local alterelectric quadrupole and periodic moir\'e model. (a) Axial basis quadrupole $Q_{z,i}\tau_z\!\sim\!x^2-y^2$. (b) Diagonal basis quadrupole $Q_{x,i}\tau_x\!\sim\!2xy$. (c) Two-component local order parameter $Q_i=Q_{z,i}+iQ_{x,i}=|Q_i|e^{i2\phi_i}$. (d) Real-space representation of $\hat Q_i=Q_{z,i}\tau_z+Q_{x,i}\tau_x$ as an electronic ellipse with orientation $\phi_i$. (e) Effective periodic moir\'e supercell in which slowly varying bilayer registry is encoded as a quadrupolar scaffold with wave vectors $q_x$ and $q_y$. (f) Competing axial ($Q_z$-dominated) and diagonal ($Q_x$-dominated) sectors in internal quadrupole space.
}
    \label{fig1}
\end{figure*}

Alterelectricity is distinct from conventional ferroelectricity and must be treated as a separate class of ordering. In a conventional ferroelectric, the primary order parameter is a polar vector and switching reverses the electric dipole moment. In an AE state, by contrast, the defining transformation changes the anisotropy of the electronic structure while leaving the scalar charge density essentially compensated. In a two-orbital description, the local low-energy space is represented by an orbital pseudospin, and the two independent traceless onsite deformations are generated by the Pauli matrices $\tau_z$ and $\tau_x$, corresponding to the axial and diagonal quadrupolar channels shown in Fig.~\ref{fig1}(a) and Fig.~\ref{fig1}(b). We therefore write the local alterelectric quadrupole as
\begin{equation}
\hat Q_i = Q_{z,i}\tau_z + Q_{x,i}\tau_x,
\end{equation}
with components
\begin{equation}
\begin{aligned}
Q_{z,i} &= \langle c_i^\dagger \tau_z c_i\rangle = |Q_i|\cos(2\phi_i), \\
Q_{x,i} &= \langle c_i^\dagger \tau_x c_i\rangle = |Q_i|\sin(2\phi_i).
\end{aligned}
\end{equation}
\(c_i\) denotes the local fermion spinor in the two-orbital basis on which \(\tau_z,\tau_x\) act. Equivalently,
\begin{equation}
Q_i = |Q_i|e^{i2\phi_i},
\qquad
\phi_i=\frac{1}{2}\arg\!\bigl(Q_{z,i}+iQ_{x,i}\bigr),
\end{equation}
so that $|Q_i|=\sqrt{Q_{z,i}^2+Q_{x,i}^2}$ measures the strength of the local orbital anisotropy and $\phi_i$ fixes the orientation of the associated electronic ellipse in real space, as illustrated in Fig.~\ref{fig1}(c) and Fig.~\ref{fig1}(d). By contrast, the scalar density $n_i=\langle c_i^\dagger \tau_0 c_i\rangle$ is independent of the quadrupolar orientation, which is precisely why distinct AE textures can remain nearly charge compensated.

The rank-two nature of the order parameter implies a double-angle transformation law under in-plane rotations,
\begin{equation}
\begin{pmatrix}
Q_z'\\
Q_x'
\end{pmatrix}
=
\begin{pmatrix}
\cos 2\theta & \sin 2\theta\\
-\sin 2\theta & \cos 2\theta
\end{pmatrix}
\begin{pmatrix}
Q_z\\
Q_x
\end{pmatrix},
\label{eq:doubleangle}
\end{equation}
up to the sign convention of the chosen real-orbital basis. On the square moir\'e structure used below, the two components therefore play the role of competing $B_{1g}$-like and $B_{2g}$-like channels. From this viewpoint, the AE doublet behaves as an internal orientational degree of freedom subject to lattice anisotropy: the problem is not only whether $|Q|$ becomes large, but also how rigidly the superlattice selects the angle in internal quadrupole space.

To isolate this orientational physics, we adopt a coarse-grained description on an effective periodic moir\'e lattice. Each site represents a local patch carrying two active orbitals, with spinor $c_i=(c_{i1},c_{i2})^T$. The local registry enters through two periodic fields, $\chi_z(\mathbf r)$ and $\chi_x(\mathbf r)$, which transform as the two components of a quadrupole. The single-particle Hamiltonian entering the self-consistency loop is
\begin{equation}
\begin{aligned}
H_{\mathrm{MF}} &= \sum_{\langle ij\rangle} c_i^\dagger T_{ij} c_j + \sum_i c_i^\dagger \Big[(u_{\mathrm m}\chi_{z,i}-KQ_{z,i})\tau_z \\
&\qquad + (u_{\mathrm m}\chi_{x,i}-KQ_{x,i})\tau_x - \mu\tau_0\Big] c_i ,
\end{aligned}
\label{eq:Hmodel}
\end{equation}
where $T_{ij}$ denotes orbital-dependent hopping, $u_{\mathrm m}$ is the moir\'e-field amplitude, and $K$ is the effective interaction scale in the quadrupolar channel. $K$ plays a role analogous to the coupling that drives an itinerant symmetry-breaking instability, now in the quadrupolar rather than spin-dipolar channel.\cite{Stoner1938,Pomeranchuk1958,Oganesyan2001,Fradkin2010} Total-energy comparisons reported below include the standard mean-field double-counting correction in the quadrupolar sector. The moir\'e fields are chosen as
\begin{equation}
\begin{aligned}
\chi_z(\mathbf r) &= \cos\theta_x - \cos\theta_y, \\
\chi_x(\mathbf r) &= \lambda \sin\theta_x \sin\theta_y
\end{aligned}
\label{eq:moirefields}
\end{equation}
with $\theta_x \equiv 2\pi m_x x / L_x$ and $\theta_y \equiv 2\pi m_y y / L_y$. The square moir\'e structure defined by Eqs.~(\ref{eq:moirefields}) is sketched in Fig.~\ref{fig1}(e). Distinct internal sectors are generated by the rotation $\chi_z+i\chi_x \rightarrow e^{i\varphi}(\chi_z+i\chi_x)$, with $\varphi=0$ and $\varphi=\pi/2$ corresponding to the axial and diagonal reference sectors summarized in Fig.~\ref{fig1}(f).

\begin{figure*}[htp]
    \centering
    \includegraphics[width=0.995\linewidth]{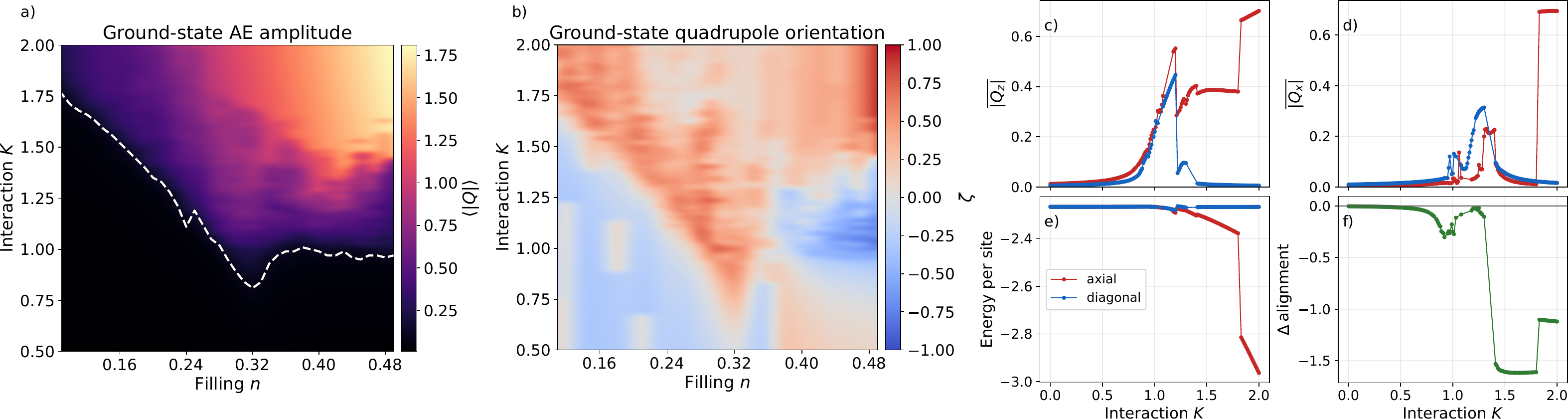}
    \caption{Self-consistent alterelectric response in the square moir\'e model. (a) Spatially averaged quadrupole amplitude $\overline{|Q|}$ across the $(n,K)$ plane; the dashed line is a guide to the crossover into the large-amplitude regime. (b) Orientation index $\zeta=(\overline{|Q_z|}-\overline{|Q_x|})/(\overline{|Q_z|}+\overline{|Q_x|})$, with $\zeta>0$ ($\zeta<0$) denoting axial (diagonal) dominance. For $n=0.30$, panel (c) shows $\overline{|Q_z|}$ and panel (d) shows $\overline{|Q_x|}$ for axial and diagonal initializations, panel (e) compares the total energies per site, and panel (f) shows the difference in normalized registry-alignment overlap between the two branches.}
    \label{fig2}
\end{figure*}

The self-consistent quadrupoles are obtained from a static site-resolved Hartree-Fock decoupling in the local quadrupolar channel~\cite{Hartree_1928,Fock1930}. Mean fields and local densities are defined in real space on the sites of the moir\'e cell, while periodic repetition is handled in reciprocal space by diagonalizing the Bloch Hamiltonian on a finite $k$ mesh. For the calculations reported here, the supercell has size $L_x\times L_y=8\times 8$, corresponding to $N=64$ moir\'e sites with two active orbitals per site, and the reciprocal-space sampling uses a $10\times 10$ mesh of the moir\'e Brillouin zone. The local quadrupoles are updated from the occupied Bloch eigenstates according to $Q_{\mu,i}=N_k^{-1}\sum_{\mathbf{k},\nu\in\mathrm{occ}}\langle \psi_{\mathbf{k}\nu}|i\rangle\tau_\mu\langle i|\psi_{\mathbf{k}\nu}\rangle$ with $\mu=z,x$, and the chemical potential is adjusted at each iteration to reproduce the target filling. This mixed real/reciprocal-space formulation retains the spatial quadrupolar texture inside the moir\'e unit cell while providing controlled convergence of the low-energy spectral response. We report the spatial averages $\overline{|Q_\mu|}=N^{-1}\sum_i |Q_{\mu,i}|$, $\overline{|Q|}=N^{-1}\sum_i |Q_i|$, together with the orientation index $\zeta=(\overline{|Q_z|}-\overline{|Q_x|})/(\overline{|Q_z|}+\overline{|Q_x|})$, and we also track a normalized registry-alignment overlap between the self-consistent quadrupolar texture and the imposed moir\'e field.

The resulting state diagram in the \((n,K)\) plane, where \(n\) denotes the filling, is shown in Fig.~2(a,b). Because the moir\'e field explicitly seeds the quadrupolar channel, the low-$K$ region is better viewed as a weak-response regime than as a sharply distinct symmetric phase. Nevertheless, Fig.~\ref{fig2}(a) displays a pronounced crossover into a large-amplitude self-consistent state, with a minimum interaction scale near $n\approx0.32$. This strong filling dependence already carries important physics: the relevant scale is not set by a featureless local interaction, but by the filling dependence of the quadrupolar susceptibility of the Bloch-periodic moir\'e background. This is equivalent to writting the effective stiffness of channel $\mu$ as $r_\mu(n)\approx 1-K\chi^{(0)}_{\mu\mu}(n)$, so the nonuniform crossover line maps the filling dependence of the underlying susceptibility. The accompanying orientation map in Fig.~\ref{fig2}(b) shows that this susceptibility is itself anisotropic in internal AE space, since over most of the large-amplitude regime the axial channel softens first, whereas a narrower diagonal-favored wedge survives close to onset at larger filling. The phase diagram therefore encodes two distinct pieces of information --- where the quadrupole becomes self-amplifying, and which internal sector is selected once it does.

A representative interaction scan at $n=0.30$ clarifies how this selection is built up. At small $K$, the axial and diagonal initializations in Fig.~\ref{fig2}(c) and Fig.~\ref{fig2}(d) yield small order parameters and nearly identical energies in Fig.~\ref{fig2}(e), meaning that the electronic response mainly tracks the imposed structural field. In this regime the AE texture is better viewed as a susceptibility imprint of the moir\'e structure than as a rigidly formed ordered state. As $K$ increases toward the crossover, both branches grow rapidly, but not monotonically. This nonmonotonicity is significant: it signals a shallow orientational energy landscape in which the amplitude has already become appreciable while the internal angle remains soft. The coexistence of sizable $Q_z$ and $Q_x$ responses, together with the still modest energy splitting, is the finite-cell mean-field signature of a nearby rotational manifold in quadrupole space rather than of two unrelated solutions.

Once $K$ is increased further, the soft manifold collapses into a clearly selected sector. The axial branch retains the larger $\overline{|Q_z|}$ [Fig.~\ref{fig2}(c)], whereas the diagonal branch loses order-parameter weight even if it can still be converged as a metastable solution. The total-energy difference in Fig.~\ref{fig2}(e) and the alignment metric in Fig.~\ref{fig2}(f) then become mutually reinforcing: the branch that best matches the structural quadrupolar field is also the lower-energy one. 
The square structure therefore acts as an orientational anisotropy in internal AE space, converting a nearly degenerate quadrupolar doublet into a selected axial ground state. From the standpoint of functionality, Fig.~\ref{fig2} separates a soft-response window, where modest perturbations should most efficiently reorient the AE texture, from a deeper ordered regime where the selected sector is thermodynamically and spectroscopically robust.

\begin{figure}[htp]
    \centering
    \includegraphics[width=0.995\linewidth]{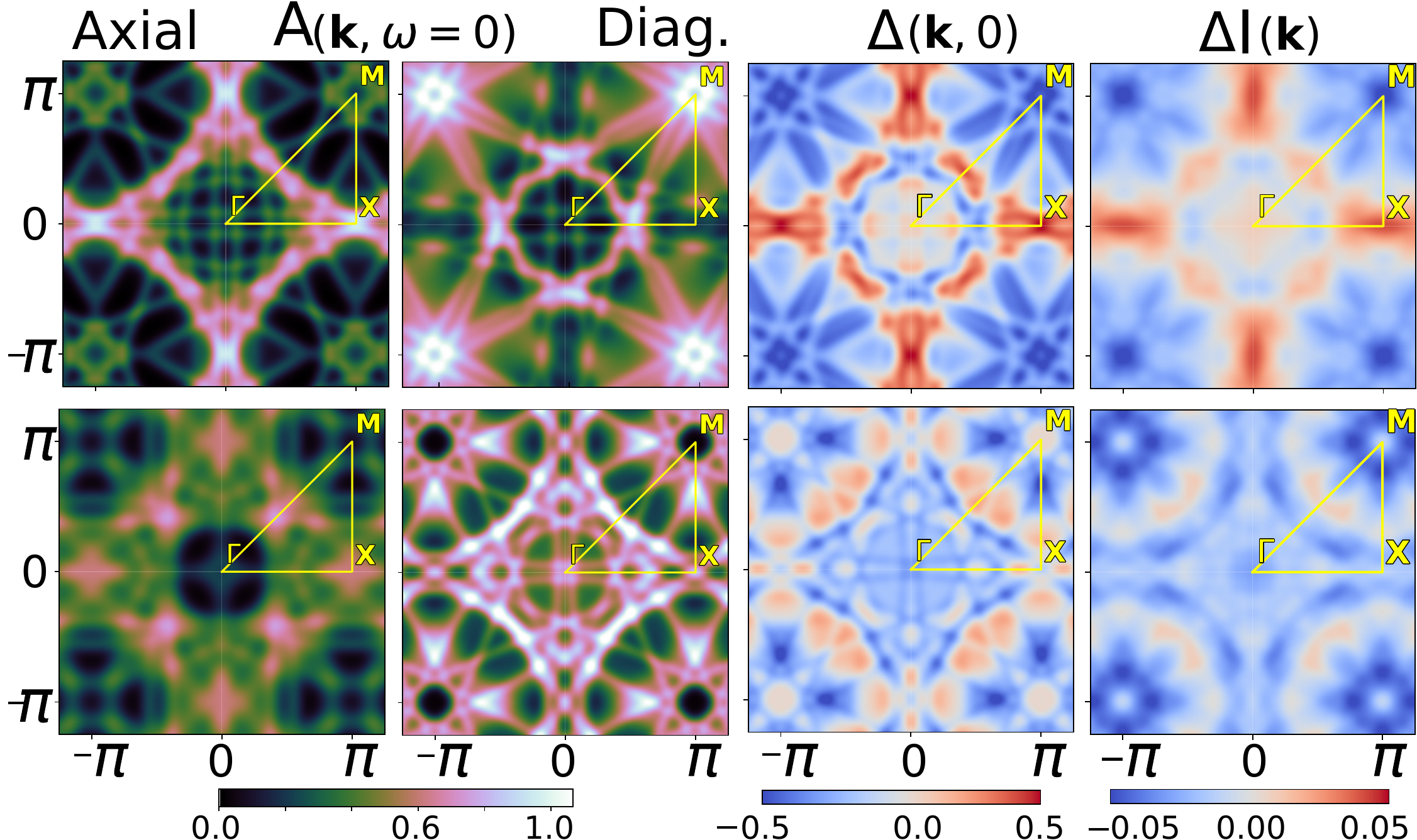}
    \caption{Momentum-resolved spectral response of the competing AE sectors at $n=0.30$. The upper row corresponds to the near-onset regime $K=1.0$, and the lower row to the stronger-coupling regime $K=1.5$. In each row, the first and second panels show $A_{\mathrm{ax}}(\mathbf{k},0)$ and $A_{\mathrm{diag}}(\mathbf{k},0)$ for the axial and diagonal branches, the third panel shows $\Delta A(\mathbf{k},0)$, and the fourth panel shows $\Delta I_{\Omega}(\mathbf{k})$. Yellow lines indicate the irreducible wedge of the moir\'e Brillouin zone.}
    \label{fig3}
\end{figure}

Throughout this evolution the state remains compensated in the sense relevant to alterelectricity: the leading order parameter is a traceless onsite quadrupole rather than a uniform dipole. The physically important distinction between branches lies instead in the anisotropy they imprint on the low-energy spectral function. This makes momentum-resolved spectroscopy the natural diagnostic of the selected sector.

The one-particle fingerprint of orientational selection is encoded in the momentum-resolved spectral function,
\begin{equation}
A(\mathbf{k},\omega)= -\frac{1}{\pi}\,\mathrm{Im}\,\mathrm{Tr}\bigl[\omega+i\eta-H_{\mathrm{MF}}(\mathbf{k})\bigr]^{-1},
\end{equation}
evaluated for the converged self-consistent solutions, together with the low-energy integral
$I_{\Omega}(\mathbf{k}) = \int_{-\Omega}^{\Omega} d\omega\,A(\mathbf{k},\omega),$
where the same cutoff $\Omega$ is used throughout the numerical analysis. Figure~\ref{fig3} compares the axial and diagonal branches through $A_{\mathrm{ax}}(\mathbf{k},0)$, $A_{\mathrm{diag}}(\mathbf{k},0)$, their difference $\Delta A(\mathbf{k},0)$, and the corresponding integrated difference $\Delta I_{\Omega}(\mathbf{k})=I_{\Omega,\mathrm{ax}}(\mathbf{k})-I_{\Omega,\mathrm{diag}}(\mathbf{k})$.

In the near-onset regime $K=1.0$, the axial and diagonal branches carry comparable integrated low-energy weight, but the first and second panels of the upper row of Fig.~\ref{fig3} show that they distribute it differently across momentum space. The axial solution reinforces a cross-like zero-energy structure through the zone center, whereas the diagonal solution transfers more weight toward off-axis lobes and corner-adjacent features. The difference map in the third panel of the upper row consequently resolves orientational selection before the total-energy splitting becomes large, and the fourth panel of the upper row shows that the contrast already extends over a finite low-energy window. In the weak-response regime, the most sensitive fingerprint of sector selection is therefore geometric: the two branches organize nearly the same low-energy phase space into different momentum-space patterns.

At $K=1.5$, the distinction is no longer a subtle rearrangement of comparable patterns. The first and second panels of the lower row of Fig.~\ref{fig3} show that the selected axial branch more strongly suppresses zero-energy spectral weight around $\Gamma$ and along extended portions of the moir\'e Brillouin zone, while the diagonal branch retains broader low-energy intensity. This is encoded quantitatively by the third panel of the lower row and, even more clearly, by the fourth panel, where $\Delta I_{\Omega}(\mathbf{k})$ is negative over much of the zone. The branch favored by the total-energy comparison is therefore also the one that depletes low-energy spectral phase space most efficiently. In this sense, the spectroscopy tracks the same electronic reconstruction that stabilizes the selected sector.

This connection is conceptually important. The internal quadrupolar orientation is not a hidden label attached to an otherwise fixed ordered state. It determines how the self-consistent orbital anisotropy is embedded into the moir\'e spectral function. Near the crossover, momentum-resolved probes would detect the impending sector selection as a redistribution pattern even when thermodynamic splittings remain small; deeper in the ordered regime, they should resolve a branch-dependent depletion of low-energy weight over extended momentum regions. The AE sector is therefore encoded simultaneously in real-space quadrupolar texture, total energy, and the momentum-space spectral response.

An explicit control coordinate is required if alterelectricity is to function as an addressable degree of freedom. We therefore introduce an internal registry phase $\alpha$ that rotates the structural field within the two-component quadrupolar space,
\begin{equation}
\chi_z^{(\alpha)}+i\chi_x^{(\alpha)} = e^{i\alpha}\,(\chi_z+i\chi_x),
\end{equation}
with $\alpha$ parameterizing the internal phase of the moir\'e field rather than a rigid real-space rotation of the lattice.

\begin{figure}[t]
    \centering
    \includegraphics[width=0.995\linewidth]{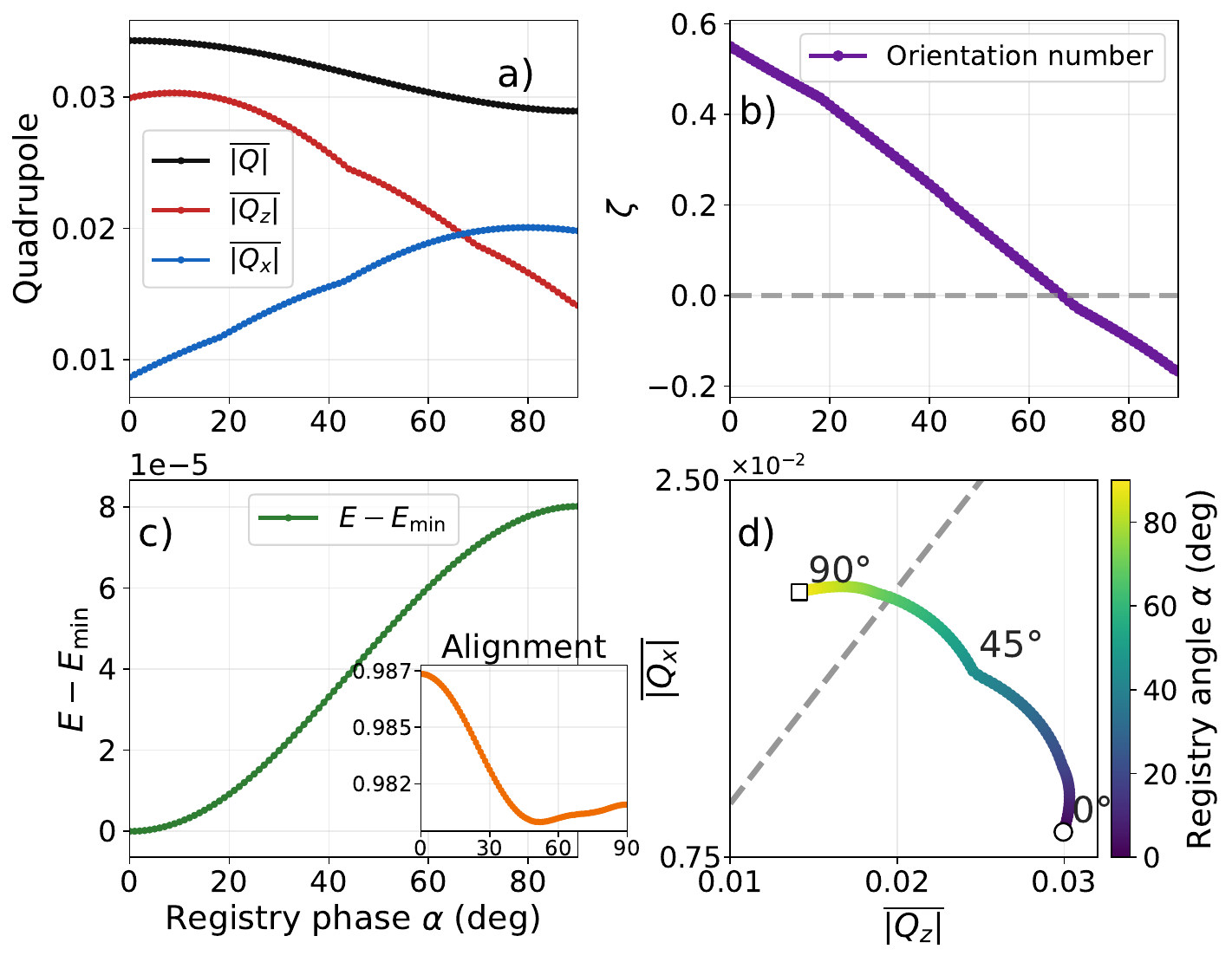}
    \caption{Continuous registry-phase steering of the alterelectric sector. (a) Spatial averages $\overline{|Q|}$, $\overline{|Q_z|}$, and $\overline{|Q_x|}$ as functions of the registry phase $\alpha$. (b) Orientation index $\zeta$, which crosses zero as the dominant quadrupolar component changes from axial to diagonal. (c) Excess energy $E-E_{\min}$ versus $\alpha$; inset: normalized registry-alignment overlap. (d) Trajectory of the self-consistent solution in the $(\overline{|Q_z|},\overline{|Q_x|})$ plane, colored by $\alpha$. The dashed line marks $\overline{|Q_z|}=\overline{|Q_x|}$.}
    \label{fig4}
\end{figure}

The sweep shown in Fig.~\ref{fig4} establishes such a control path. The total amplitude $\overline{|Q|}$ remains finite and only weakly renormalized throughout the scan [Fig.~\ref{fig4}(a)], while weight is continuously transferred from $\overline{|Q_z|}$ to $\overline{|Q_x|}$. Correspondingly, the orientation index $\zeta$ decreases monotonically and crosses zero only after a substantial registry-phase rotation [Fig.~\ref{fig4}(b)], demonstrating a continuous evolution from an axial-dominated to a diagonal-dominated texture under a single structural parameter.

That shift carries nontrivial physics. If the order parameter merely followed the imposed phase as a passive vector, equal axial and diagonal weights would be reached near $\alpha=45^\circ$ and the path in Fig.~\ref{fig4}(d) would approach a simple circular rotation. Instead, the trajectory is bowed and the equal-weight line is crossed only after the imposed phase has already strongly favored the diagonal channel. The square moir\'e scaffold therefore leaves a residual axial bias even during the sweep, and the registry phase must work against a genuine internal orientational stiffness. The crossover angle thus measures the self-consistent anisotropy of the AE manifold rather than a trivial geometrical interpolation.

Equally important, the conversion occurs without destroying the order. The excess energy varies smoothly and remains very small over the entire scan [Fig.~\ref{fig4}(c)], while the normalized registry-alignment overlap shown in the inset of Fig.~\ref{fig4}(c) changes only modestly. The control coordinate therefore rotates the order within quadrupole space much more than it suppresses its amplitude. The central functional result of this study can be summarized as: moir\'e registry supplies a continuous structural knob that connects alterelectric partner sectors through an adiabatic path. In a material setting, such a coordinate could be realized by controlled interlayer translation, strain-assisted reconstruction, or piezoelectrically driven registry shifts in orbital-active bilayers.~\cite{Alterelectricity2026,Carr2020Methods,Nuckolls2024Moire}

To conclude, we have shown that a Bloch-periodic moir\'e environment can both bias and steer a compensated alterelectric quadrupole. At the reference registry phase, the square moir\'e field lifts the near-degeneracy between axial and diagonal sectors and selects an axial-dominated ground state over most of the large-amplitude regime. The state diagram separates a soft window of enhanced orientational susceptibility from a deeper ordered regime in which the selected sector is robust. The same branch selection is encoded directly in the momentum-resolved spectral function through a characteristic redistribution and depletion of low-energy weight across the moir\'e Brillouin zone.

Introducing an internal registry phase upgrades this static selection to continuous control. The self-consistent solution traverses from $Q_z$-dominated to $Q_x$-dominated textures while retaining a finite quadrupole amplitude, establishing that moir\'e registry acts on the internal orientation of alterelectric order rather than merely on its magnitude. Because the control coordinate is structural, the most immediate material implications concern registry-tunable layered systems. The alterelectric proposals of sliding bilayers identify the relevant symmetry ingredients~\cite{Alterelectricity2026}, while moir\'e heterostructures supply the spatial amplification and spectroscopic access needed to convert those ingredients into a continuous control manifold.~\cite{Nuckolls2024Moire} More broadly, the present results suggest that compensated quadrupolar order should be viewed as a programmable electronic texture in layered materials. Material-specific parameterization, finite-temperature stability, disorder, and nonequilibrium driving of the registry coordinate remain natural next steps.

\section{Acknowledgements}
Work was carried out under the auspices of the U.S. DOE through the Los Alamos National Laboratory, operated by Triad National Security, LLC (Contract No. 892333218NCA000001). 

\bibliographystyle{apsrev4-2}
\setcitestyle{notesep={;}}
\bibliography{biblio.bib}

\end{document}